# Transportation Internet: Concepts, Models, and Architectures

Hui Li, Yongquan Chen, *Member, IEEE,* Keqiang Li, *Member, IEEE,* Chong Wang, Bokui Chen

*Abstract*—Disruptive changes in vehicles and transportation have been triggered by automated, connected, electrified and shared mobility. Autonomous vehicles, like Internet data packets, are transported from one address to another through the road network. The Internet has become a general network transmission paradigm, and the Energy Internet is a successful application of this paradigm to the field of energy. By introducing the Internet paradigm to the field of transportation, this paper is the first to propose the Transportation Internet. Based on the concept of the Transportation Internet, fundamental models, such as the switching, routing, and hierarchical models, are established to form basic theories; new architectures, such as transportation routers and software defined transportation, are proposed to make transportation interconnected and open; system verifications, such as prototype and simulation, are also carried out to prove feasibility and advancement. The Transportation Internet, which is of far-reaching significance in science and industry, has brought systematic breakthroughs in theory, architecture, and technology, explored innovative research directions, and provided an Internet-like solution for the new generation of transportation.

## I. INTRODUCTION

IN recent years, technologies such as cloud computing, fifth generation (5G), artificial intelligence (AI), autonomous driving, and vehicle-to-Everything (V2X) are reshaping vehicles and transportation. Automated, connected, electrified, shared (ACES) mobility is leading the transformation of transportation to the next paradigm [1]. Experts predict that by 2040, approximately 75% of vehicles will be autonomous, connected, and electric, potentially with shared ownership [2]. Researches on connected and automated vehicles (CAV) [3], connected and automated driving (CAD) [4], and connected and automated transportation (CAT) [5] have been conducted. A new generation of transportation system urgently needs to be established [6].

After matured in desktop computers, the Internet has greatly expanded within the field of information, producing the Mobile Internet, Internet of Things (IoT), Industrial Internet, Internet of Vehicles (IoV), Physical Internet, and so on [7]–[11].

The Internet further expands beyond the field of information. The Internet paradigm was introduced into the field of energy, and the Energy Internet was proposed [12]. In the Energy Internet, the energy network is similar to the communication network, the energy flow is similar to the information flow, and the energy router is similar to the information router, thus forming an Internet-like energy infrastructure [13].

Inspired by the Energy Internet, we introduce this paradigm to the field of transportation. In our proposed system, the road network is similar to the communication network, the traffic flow is similar to the information flow, and the autonomous vehicle is similar to the data packet. The Transportation Internet is proposed for the first time and offers a systematic solution for the new generation of transportation.

In this paper, we establish the fundamental theories such as the transportation switch model, routing model, and hierarchical model for the Transportation Internet; propose the transportation router, and present the architecture of the transportation router; further build SDT by separating the control plane of the transportation router to make transportation programmable and open.

The rest of this paper is organized as follows. Section II describes the related research work. Section III introduces the fundamental concepts. Section IV introduces the fundamental models. Section V presents the architectures of transportation routers and SDT. Section VI introduces evaluations. Section VII draws the conclusions.

## II. RELATED WORK

Intelligent transportation systems (ITS) were formally produced in the 1990s, which include Advanced Traffic Management Systems (ATMS), Advanced Travelers Information Systems (ATIS), Advanced Vehicle Control Systems (AVCS), and other subsystems [14], [15], and provide a series of chimney applications, such as traffic signal control, electronic police, variable message signs, electronic toll collection, cooperative ITS (C-ITS, also known as V2X), etc. [16]. Recently, some applications of ITS have been combined with emerging technologies, including cloud computing, IoT, big data, and AI, etc. [17]–[19] Some new system schemes have also been proposed. ITS and cyber-physical system (T-CPS) are combined, through computing, communication, control (3C) technology to integrate information components and physical components to realize interaction and feedback between the cyber system and physical system [20], [21]. The parallel transportation system (PTS), which is based on artificial society, computing experiment, and parallel execution (ACP) theory, creates an artificial transportation system equivalent to a physical transportation system in the computer. Based on the PTS, the evolution process of the transportation system is studied from the perspective of a parallel world [22]–[24].

Due to the limitations of early technologies, ITS vertically formed chimney architectures, which led to problems such as system closure, data fragmentation, difficulty of intercommunication, etc. Some applications of ITS, that are primarily combined with cloud computing, big data, AI and so on, mainly a combination with a single application, have problems such as low system resource utilization, low intelligence, and lack of application ecology. New transportation architectures, such as T-CPS and PTS, have not considered these problems.

As a natural extension of ITS, CAV is a major development based on communication and computing technologies [25]. Path planning is an important research field of CAV, including technologies such as path planning for automated driving [26] and Vehicle Routing Problem (VRP) for logistics [27]. At present, global planning only supports a small number of vehicles with a static number and goal, and local planning only supports a small number of determined dynamic factors. Moreover, planning algorithms take a long time. So, it is difficult to adapt to the real-time dynamic planning requirements of the large-scale autonomous driving.

Intelligent connected vehicles (ICV) further offer a systematic solution [28] and propose a cloud control system (CCS) combined with V2X. CCS interconnects pedestrians, vehicles, roads, and clouds for integrated perception, decision making and control [29]. CCS is essentially a vehicle-road collaboration system constructed by ICV and C-ITS. As CITS is one of the chimneys of ITS, the chimney architecture of ITS is not changed.

The Internet is a communication network based on packets switching technology. The Internet Protocol (IP) constitute the cornerstone of the Internet for routing requirements. Routers achieve an automatic information transmission network based on IP protocol, creating a large-scale information infrastructure and an open application ecology [30]. Traditional routers have many problems, such as complex structure, low flexibility, and difficult management. The architecture of software defined network (SDN) separates the control plane of the router from the underlying switch (data plane) by introducing OpenFlow interface, which makes network programmable, and simplifies policies implementation and network management [31]–[33].

As mentioned in Section I, the Internet has expanded to other information fields, such as the Mobile Internet, the Industrial Internet, IoT, IoV and PI are produced [11]. IoV is regarded as the combination of IoT and V2X [34]. Cellular V2X (C-V2X) is an important IoV enabling technology for autonomous driving and ITS [35], [36]. SDN technology provides IoV with the capability of global network optimization and efficient network resource coordination [37], and can be further combined with AI, edge computing, and other technologies [38]. PI has developed to be a global logistics system based on IoT. Each container has a unique identification and specification for easy end-to-end identification and tracking [39]. Therefore, IoV and PI primarily provide communication solutions and focus on local transportation problems.

In general, the current transportation approaches, which are mainly based on the chimney architecture of ITS, have difficulty meeting the demands of disruptive changes for the new generation of transportation.

## III. TRANSPORTATION INTERNET CONCEPTS

The Internet has advanced in theories, technologies, architectures, and applications, and has become a general paradigm for large-scale network transmission. Therefore, the Internet paradigm not only supports the transmission of information but also the transmission of energy and matter. For the purpose of distinction, the existing Internet that transmits information is hereinafter referred to as the Information Internet.

We introduce the Internet paradigm to the field of transportation and propose the Transportation Internet, which is used as the typical form of matter transmission. The Transportation Internet, together with the Information Internet and Energy Internet, forms three pan-Internet systems (Fig. 1).

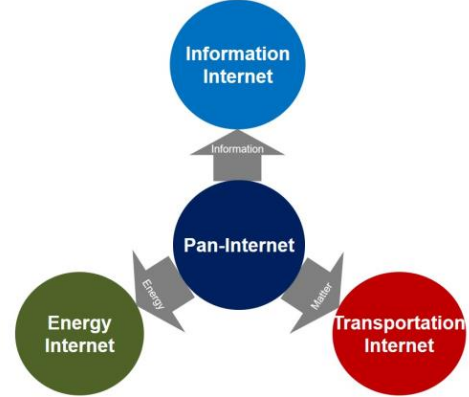

Fig. 1: Three pan-Internet systems

A communication network system consists of basic elements such as terminal equipment, transmission lines, and switching equipment. We establish corresponding concepts for the Transportation Internet.

### A. Transportation Terminal

The transportation terminal corresponds to the network terminal (host) of the Information Internet. The transportation terminal is the abstraction of all types of transportation locations for mobility, such as parking lots and stops. The transportation terminal is described by the transportation address. Each transportation terminal is assigned a logical address to shield the difference of physical (geographical) address and provide a routing function, which has the same status in the Transportation Internet.

### B. Transportation Line

The transportation line corresponds to the network line of the Information Internet. The transportation line is the abstraction of all types of roads. One road contains several lanes.
A lane is abstracted as a transportation channel. The transportation network is formed by interconnected transmission lines, and the transportation node is formed by the junction of the transmission lines. Typical transportation nodes include intersections, inbounds, outbounds, bridges, and tunnels.

### C. Transportation Router

The transportation switching equipment is deployed on the transportation node and is responsible for the dispatch of traffic signals and nearby vehicles. According to the complexity of traffic dispatch, the transportation switching equipment is divided into the transportation router (e.g., with signal control) and transportation switch (e.g., without signal control). For the convenience of description, we collectively refer to transportation switching equipment as transportation routers.

*D. Transportation Vehicle*

The transportation vehicle corresponds to the data packet of the Information Internet. The transportation vehicle includes all types of means passing through the transportation network, including pedestrians according to the situation. Transportation vehicles carry necessary information such as originating and destination addresses, priority, and so on. However, transportation vehicles also have different characteristics than data packets. Transportation vehicles have the capability of autonomous driving, which cannot be copied, compressed or discarded, and may malfunction or be damaged. Transportation vehicles also have license plates, which can be identified and tracked.

The comparison of the concepts of the three pan-Internet systems is shown in Table I.

TABLE I: Comparison of the three pan-Internet systems

| Type | Information Internet | Energy Internet | Transportation Internet |
|---|---|---|---|
| Terminal Equipment | Network terminal | Energy terminal | Transportation terminal |
| Transmission Line | Network line | Energy line | Transportation line |
| Switching Equipment | Network router/switch | Energy router/switch | Transportation router/switch |
| Transmission Content | Data packet | Electric power | Transportation vehicle |

## IV. TRANSPORTATION INTERNET MODELS

Routing is the core requirement of the Information Internet, which transport information from one address to another. The routing process has several aspects such as collecting network status, calculating feasible routes, and forwarding data packets.

Similarly, routing is the core requirement of the Transportation Internet, which transport vehicles from one address to another. We establish the network model of the Transportation Internet to describe transportation routing requirements(Fig. 2), and further build the switching model (for dispatching vehicles), routing model (for calculating feasible routes), and hierarchical model (for managing network status) to provide routing capabilities based on transportation routers for the large-scale network.

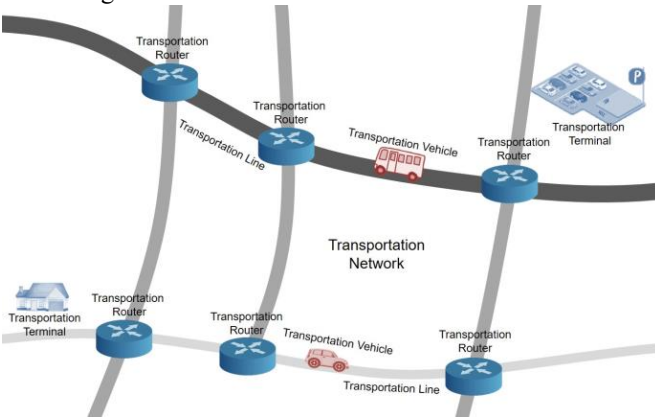

Fig. 2: Network model of the Transportation Internet

The Transportation Internet provides routing services based on the best-effort model, that is, the routing problem under unconstrained conditions. The transportation router fairly selects routes for all vehicles according to the shortest path algorithm. For special circumstances, such as ambulances and fire trucks, the Transportation Internet must also introduce a quality of service (QoS) model to provide deterministic services [40]. The QoS model is a routing problem with constraints. Therefore, the best-effort model can be regarded as a special case of the QoS model. We establish a unified model for routing requirements.

Suppose that a directed graph $G = (V,E)$ is used to represent the transportation network, where $V$ represents the set of all nodes (i.e., transportation routers) in the topology; $E$ represents the set of all edges (i.e., transportation lines, also known as links) of the topology. The routing problem of the Transportation Internet can be described as searching for a path p, which satisfies the specified constraints and optimization goals, from $s$ (source node) to $t$ (destination node). Let $p = v_s v_{s+1}...v_{t-1} v_t$.

Suppose constraints are described by $N$ constraint metrics, $w_k(v)$ denotes the $k$-th constraint metric attached to the node $v \in p$, $w_k(e)$ denotes the $k$-th constraint metric attached to the edge $e \in p$, and $w_k(p)$ denotes the $k$-th constraint metric attached to the path $p$. Based on the different characteristics of transportation metrics, there are three expressions for $w_k(p)$ as follows:

(i) If $w_k(p) = \Sigma_{v \in p} w_k(v) + \Sigma_{e \in p} w_k(e)$, then the transportation metric is additive, such as distance, time, and cost.

(ii) If $w_k(p) = \Pi_{v \in p} w_k(v) \cdot \Pi_{e \in p} w_k(e)$ then the transportation metric is multiplicative, such as accident rate.

(iii) If $w_k(p)=\{\max\{w_k(v)_{v \in p}\}, \min\{w_k(v)_{v \in p}\}, \max\{w_k(e)_{e \in p}\}, \min\{w_k(e)_{e \in p}\}\}$, then the transportation metric is concave, such as the overweight limit and overspeed limit.

Therefore, routing requirements can be expressed as the following two basic problems and their combinations:

(i) Routing optimization problem: take the metric $w_k(p)$ as the objective function

$$\min w_k(p) \quad or \quad \max w_k(p) \quad (1)$$

(ii) Routing constraint problem: take the constraint $\delta_k$ of the metric $w_k(P)$ as the constraint:

$$w_k(p) \leq \delta_k \quad or \quad w_k(p) \geq \delta_k \quad (2)$$

Based on the network model, we further establish the following three specific models for routing requirements.

*A. Model of Transportation Switching*

In the Information Internet, packet-switching technology based on time-division multiplexing is used to transmit data packets. A typical packet switching fabric is a switching matrix in which any input port is connected to any output port through a high-speed crossbar switch circuit. The transportation node can be equivalent to a packet-switching fabric based on timedivision multiplexing. A road intersection is a typical example, in which each lane corresponds to an input port or output port, forming a switching matrix connected by any input port and any output port. The switching model of a typical intersection (two-way, four-lane) is illustrated in Fig. 3. The dispatching function of the transportation router controls the access of the switching fabric according to certain rules to achieve efficient traffic.

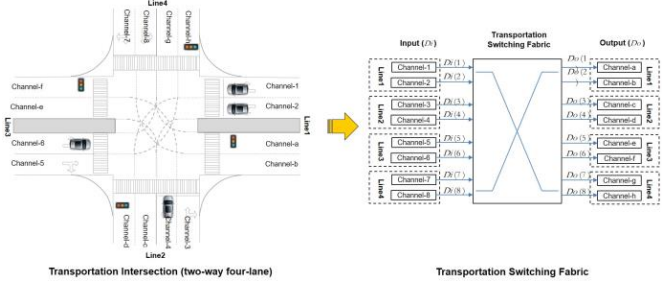

Fig. 3: Switching model of intersection (two-way, four-lane intersection)

Queuing is introduced for the input and output ports. Based on queuing theory, the appropriate queuing algorithm is selected according to destination addresses of vehicles, the distribution of arrivals, the distribution of the intersection processing time, and the number of service desks. According to whether there is signal control at the intersection, the algorithm can be divided into periodic traffic queuing algorithm and random traffic queuing algorithm. The periodic traffic queuing algorithm can be modeled by referring to the M/D/1 and M/Dx/1 models, while the random traffic queuing algorithm can be modeled by referring to the M/M/1 and M/G/1 models.

Switching is introduced for connection matching between the input and output ports. Some competition results of input and output ports give way to other input and output ports to form a port-matching connection set. The traffic conflicts in intersecting areas include cross conflicts, merge conflicts, and diversion conflicts. The available switching algorithms are the round-robin algorithm, longest queue first algorithm, maximum weighted matching algorithm, etc.

Suppose that a bipartite graph $G = (D,C)$ represents the switching fabric where $D$ represents the set of all input and output ports and $C$ represents the set of all connections between input and output ports. Suppose that $v$ denotes an internal path formed by an input port and output port, and the connection between them, $w_k(v)$ denotes the $k$-th constraint metric attached to the path $v$, $w_k(d)$ denotes the $k$-th constraint metric attached to the nodes $d \in D$, and $w_k(c)$ denotes the $k$-th constraint metric attached to the edges $c \in C$.

According to the type of metric, $w_k(v)$ is calculated with $w_k(d)$ and $w_k(c)$. The routing problem of the switching fabric can be described as follows:

(i) Routing optimization problem of the switching fabric: take the metric $w_k(v)$ as the optimization goal

$$\min w_k(v) \quad or \quad \max w_k(v) \qquad (3)$$

(ii) Routing constraint problem of the switching fabric: take the constraint $\delta_k$ of the metric $w_k(v)$ as the constraint target

$$w_k(v) \leq \delta_k \quad or \quad w_k(v) \geq \delta_k \qquad (4)$$

## B. Model of Transportation Routing

In the Information Internet, the router collects and continuously updates the status of the network, and then calculates feasible routes based on the collected network status.

The transportation infrastructure is the auxiliary equipment deployed on the roadside, including cameras, radars, signal lights, speed limit signs, guidance screens, etc. The transportation infrastructure is controlled by the transportation routing function of the nearby transportation router as peripherals. The transportation vehicle is controlled by the transportation routing function of the nearby transportation router through wireless communication such as 5G.

Compared with traditional map-based navigation technology, routing technology has dynamic, real-time, and largescale capabilities. Based on the transportation switching fabric, a transportation routing model is formed to replace the existing chimney model (Fig. 4), which includes the following two aspects:

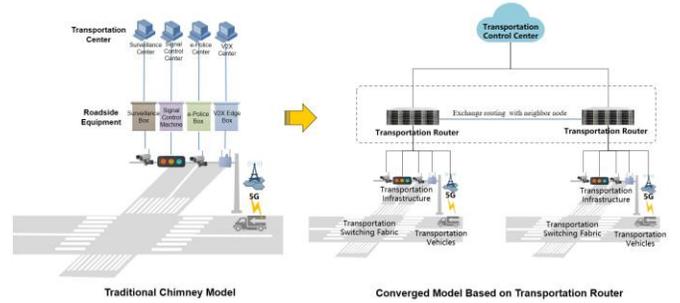

Fig. 4: Model of transportation routing

(i) Collecting network status: The transportation routing function collects traffic status (such as vehicle density, traffic incidents, etc.) through roadside peripherals (such as signal lights, cameras, radars, etc.) and collects the status (such as vehicle location, speed, etc.) of connected vehicles passing nearby. The transportation routing function establishes links with neighboring routers through routing protocols, collects the routing information of other routers, and publishes its own routing status to others.

(ii) Computing feasible routes: The transportation routing function generates the network topology based on the collected network status and calculates feasible routes according to the transportation routing algorithm. To solve the problem quickly, the network topology is limited to a certain range. Typical routing algorithms usually use the Lagrangian algorithm, branching algorithm, Bellman-Ford algorithm, Dijkstra algorithm, Floyd-Warshall algorithm, and other algorithms. The routing function sends the feasible route to the dispatching function to dispatch the transportation switching fabric.

Suppose that the directed graph $G_i = (V_i, E_i)$ represents the transportation network topology in an area, where $V_i$ represents the set of all nodes of the topology structure, and $E_i$ represents the set of all edges of the topology structure. Let $p_i$ denote the path selected in the area, $w_k(p_i)$ represent the $k$-th constraint metric on $p_i$, $w_k(v_i)$ represent the $k$-th constraint metric on the node $v_i \in p_i$, and $w_k(e_i)$ denote the $k$-th constraint metric on the edge $e_i \in p_i$.

According to the type of metric, $w_k(p_i)$ is calculated with $w_k(v_i)$ and $w_k(e_i)$. The routing problem in the $G_i$ area can be described as follows:

(i) Routing optimization problem of the $G_i$ area: take the metric $w_k(p_i)$ as the optimization target

$$\min w_k(p_i) \quad or \quad \max w_k(p_i) \tag{5}$$

(ii) Routing constraint problem of the $G_i$ area: take the constraint $delta_k$ of the metric $w_k(p_i)$ as the constraint target

$$w_k(p_i) \leq \delta_k \quad or \quad w_k(p_i) \geq \delta_k \tag{6}$$

*C. Model of Transportation Hierarchy*

For a large network, it is very difficult for a router to collect the status from the entire network. The Information Internet generally divides the entire network into multiple autonomous systems (AS), and combines the Tiers model, Transit-Stub model, and other models to achieve hierarchical routing management [41], [42].

The transportation network naturally has the characteristics of area division and classification. We divide several transportation routers that are connected to each other in the same scope and are under the control of the same administrative agency into one area, which is similarly called the transportation autonomous system (TAS). The Transportation Internet is logically composed of multiple interconnected TAS. Furthermore, based on the geographical distribution and road classification, TAS is similarly divided into stub TAS (responsible for terminal access) and transit TAS (responsible for transit). We further describe the Transportation Internet as a three-tier structure (Fig. 5).

(i) Transportation Local Area Network (LAN): This network is composed of low-level roads (e.g., secondary roads, branch roads) connected to each other in the same geographic area, and typically includes stub TAS.

(ii) Transportation Metropolitan Area Network (MAN): This network is composed of express roads (e.g., urban expressways, urban arterial roads) connecting several LAN, and typically includes transit TAS.

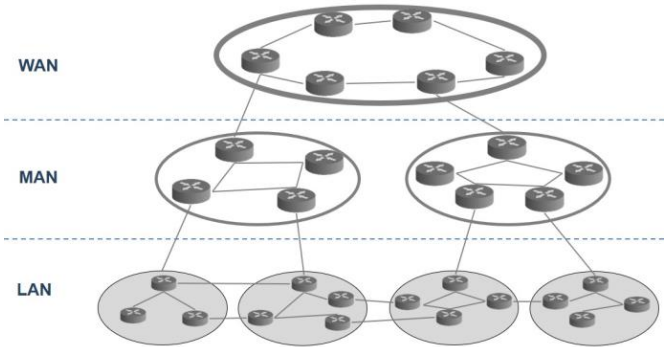

Fig. 5: Model of transportation hierarchy

(iii) Transportation Wide Area Network (WAN): This network is composed of high-speed arterial roads (e.g., highways, intercity arterial roads) connecting several LAN and MAN, and typically includes transit TAS.

Suppose that a directed graph $H = (G_i, E_e)$ represents the entire transportation network topology where $G_i$ is the set of all areas that constitute the network and $E_e$ is the set of external links connecting each area. Suppose that $p$ is the path selected in the entire network $H$, $w_k(p)$ represents the $k$-th constraint metric attached to $p$, $p_i$ is the selected intra-domain path, $w_k(p_i)$ represents the $k$-th constraint metric attached to the path $p_i$; $p_e$ is the selected inter-domain path, $w_k(p_e)$ represents the $k$-th constraint metric attached to the path $p_e$. $p = p_i \cup p_e$ and $p_i \cap p_e = \emptyset$. According to the type of metric, $w_k(p)$ is calculated with $w_k(p_i)$ and $w_k(p_e)$. The routing problem of the network $H$ can be described as follows:

(i) Routing optimization problem of the entire network $H$: take the metric $w_k(p)$ as the optimization target

$$\min w_k(p) \quad or \quad \max w_k(p) \tag{7}$$

(ii) Routing constraint problem of the entire network $H$: take the constraint $\delta_k$ of the metric $w_k(p)$ as the constraint target

$$w_k(p) \leq \delta_k \quad or \quad w_k(p) \geq \delta_k \tag{8}$$

V. TRANSPORTATION INTERNET ARCHITECTURES

*A. Architecture of Transportation Router*

Refer to the architecture of information router, based on the transportation switching and routing model, the architecture of transportation router includes the following two parts (Fig. 6).

*1) Transportation routing engine:* The routing engine provides the control plane of the transportation router based on the routing function. It builds and updates the routing database by dynamically collecting local routing information and exchanging neighbor routing information; then, it calculates the control policies based on routing algorithms, and issues the control policies to the dispatching engine for execution.

*2) Transportation dispatching engine:* The dispatching engine provides the transport plane of the transportation router based on the dispatching function. It generates dispatching rules using the control policies issued by the routing engine and executes the dispatching rules on transportation signals and vehicles based on the transportation switching fabric.

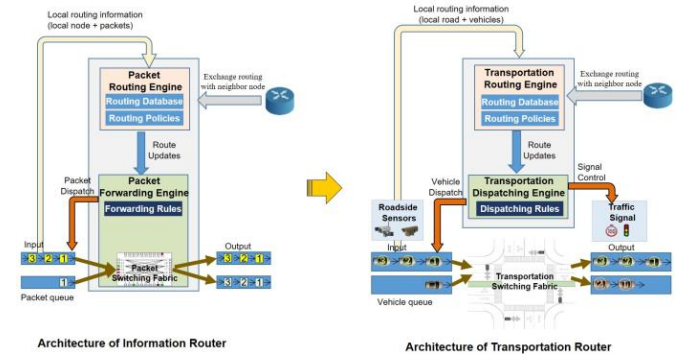

Fig. 6: Architecture of transportation router

*B. Architecture of Software Defined Transportation (SDT)*

Software definition has become a widely used network architecture paradigm. Typical cases include SDN, softswitch, etc. SDT is proposed based on the transportation router architecture. The general principle is illustrated in Fig. 7:

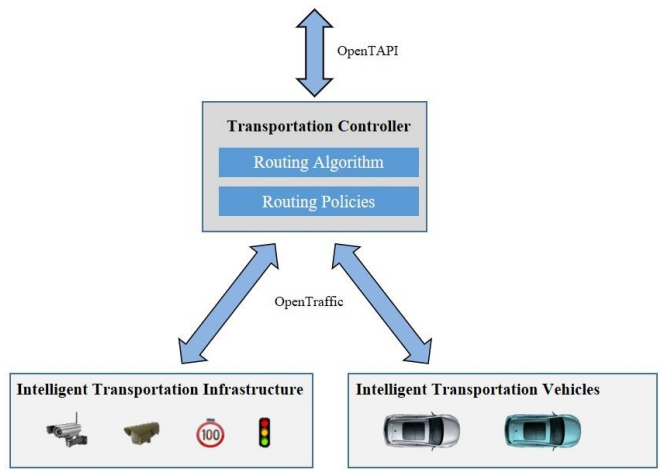

Fig. 7: Model of SDT

(i) By introducing the OpenTraffic and OpenTAPI open interfaces, the control layer (routing engine) and dispatching layer (dispatching engine) of the transportation routers are separated, and open services are provided to the outside.

(ii) A transportation controller is formed based on the separated control layer.

(iii) By integrating the separated dispatching engines with transportation infrastructure and transportation vehicles, intelligent transportation infrastructure and intelligent transportation vehicles are formed.

To facilitate the discussion, we divide SDT into the following two parts.

*1) Software Defined Transportation Infrastructure (SDT-I):* Transportation infrastructure includes the signal machine, electronic police, traffic guidance, electronic toll collection (ETC), energy charging control, and other ITS applications. Considering the similarity of infrastructure architectures, we use the signal machine as an example to research.

Through the OpenTraffic and OpenTAPI interfaces, the signal machine is separated into a signal controller and an intelligent signal light(Fig. 8). The signal controller is responsible for dynamically generating signal control policies based on external conditions and sending policies to the intelligent signal light. The intelligent signal light reports the status to the signal controller and receives policies to display the light color. The policies execute the operation of displaying the light color of the signal light by specifying the road, lane, time, signal cycles, and signal phases. These policies form "flow tables" for signal control.

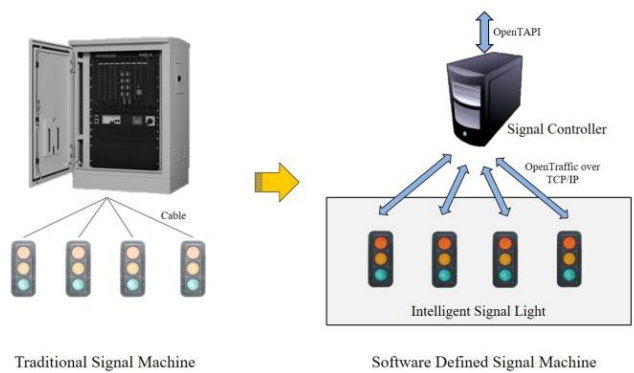

Fig. 8: Decoupling model of signal machine

Other transportation infrastructure is similarly decoupled into infrastructure controllers and intelligent infrastructure.

*2) Software Defined Transportation Vehicle (SDT-V):* Transportation Vehicles includes automobiles, pedestrians, etc. We consider the example of CAV to study SDT-V, that is, software defined vehicles (SDV). SDV includes two aspects: vehicle softwarization and separation of vehicle control. The former provides an on-board operating system for vehicle. This paper mainly discusses the latter, which separates the CAV into the intelligent vehicle and the vehicle controller (Fig. 9).

The vehicle controller generates control policies based on the external conditions and sends them to the intelligent vehicle. The intelligent vehicle reports the system status (such as automated driving level, position, speed, acceleration, steering signal, etc.), perception, planning, and other data to the vehicle controller in real time and receives policies issued by the vehicle controller to control the vehicle driving.

Control policies include operations on perception, planning, decision, and maps for designated roads, lanes, time, vehicle signs and other information. These control policies form "flow tables" for vehicle control. In the same way, the pedestrian controller and others are also implemented.

By combining SDT-I and SDT-V, the SDT hierarchical architecture is presented in Fig. 10.

(i) Transport Layer: This layer is mainly composed of intelligent vehicles, including pedestrians with intelligent devices. Intelligent vehicles collect status data to transmit to the transportation controllers and receive control policies from the transportation controller to execute control policies.

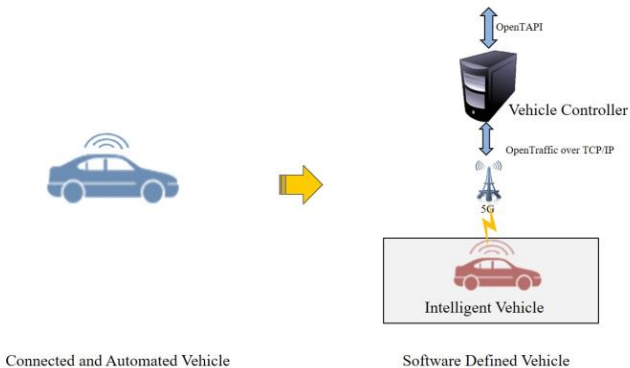

Fig. 9: Decoupling model of Connected and Automated Vehicle

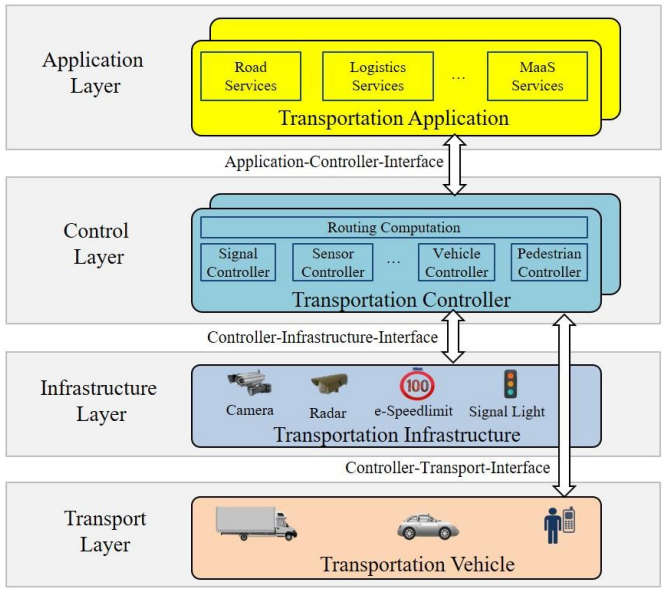

Fig. 10: Layers of SDT

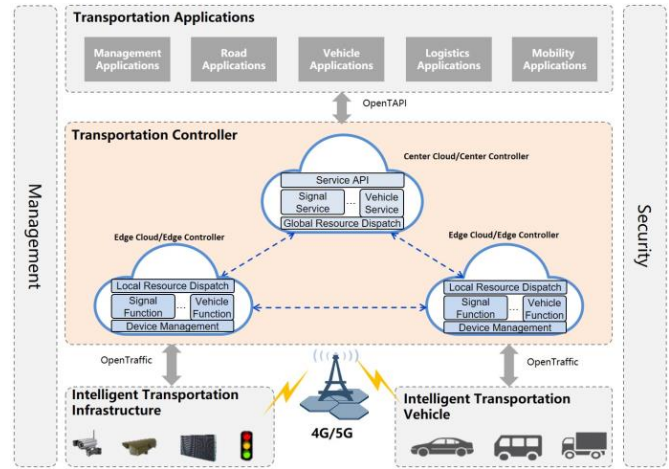

Fig. 11: Architecture of Transportation Operating System

(ii) Infrastructure Layer: This layer is mainly composed of intelligent infrastructure such as intelligent signal light, intelligent cameras, and so on. The intelligent infrastructure collects transportation status data to transmit to the transportation controller, receives control policies from the transportation controller, and executes control policies.

(iii) Control Layer: This layer is composed of a variety of transportation function controllers, including signal controllers, vehicle controllers, pedestrian controllers and so on, and performs local routing calculations for cross-function controllers. The transportation controller receives the status data transmitted by the intelligent infrastructure and intelligent vehicles through the open interface, and issues control policies to them.

(iv) Application Layer: This layer consists of various SDT applications. The transportation application is provided by calling the northbound interface of the controller. Typical applications include MaaS services, logistics services, bus services, etc., which can be collectively referred to as Transportation as a Service (TaaS).

Based on the SDT, the transportation controller can be deployed on general-purpose servers or cloud to make transportation open and programmable. For certain delay-sensitive tasks, such as autonomous driving, edge computing has become a necessary solution. The transportation controller is divided into two parts. The distributed architecture of SDT is shown in Fig. 11, and we also call it a transportation operating system (TOS).

(i) The edge controller provides local real-time functions and services for the intelligent infrastructure and intelligent vehicles through OpenTraffic API, e.g., lane-level planning, holographic intersections.

(ii) The central controller is responsible for the control of global transportation resources and provides resource coordination and dispatch, as well as open services.

## VI. TRANSPORTATION INTERNET VERIFICATIONS

To evaluate the advancement and feasibility of the Transportation Internet, we designed a prototype system and a simulation system to verify the scenarios of semi-closed campus and open roads.

### A. System Experiments, Campus of CUHK, Shenzhen

We selected a ring road on the campus of the Chinese University of Hong Kong (CUHK), Shenzhen as the verification site. The total length is approximately 1 km. There are four intersections, two of which have with signal lights. In one of the sections, it is the entrance and exit of two dormitories. During breaks, there is a large amount of student traffic, and there is a shuttle bus stop nearby, and buses pass regularly. We have designed a number of applications around the ring road, including dynamic route planning, changing lanes by consultation, perception beyond line-of-sight, cooperative merging, cooperative signal control, remote driving and valet parking (Fig. 12 and Fig. 13). In fact, because the Transportation Internet is an Internet-like platform, capabilities and applications are not limited in the prototype system.

The prototype system includes a SDT controller, two CAVs, intelligent signal lights, roadside cameras and radars, 5G networks, and so on and provides some typical applications for verification (Table II).

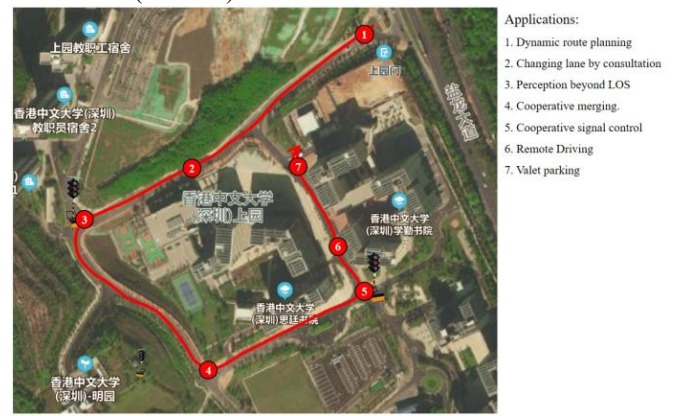

Fig. 12: Site and applications of verification

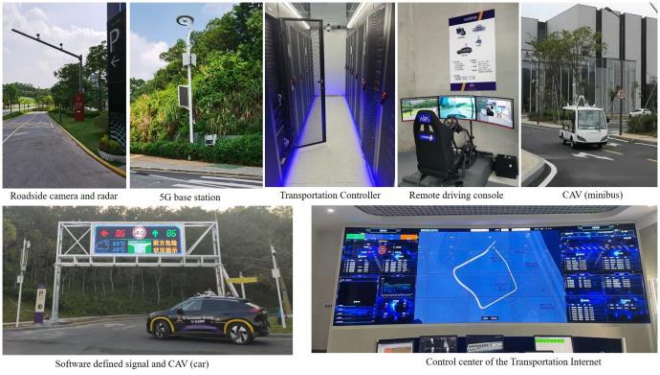

Fig. 13: Composition of the Transportation Internet verification system

There are multiple indicators to verify the Transportation Internet. We select the average vehicle speed indicator to comprehensively judge the routing capability. The CAV travels along the ring road for one loop (speed limit is 30km/h), and the average speed is calculated and compared between the system with and without SDT Controller. From the overall test results in Fig. 14, the average vehicle speed has increased by about 14% under Controller control, which proves that the Transportation Internet can improve the traffic efficiency.

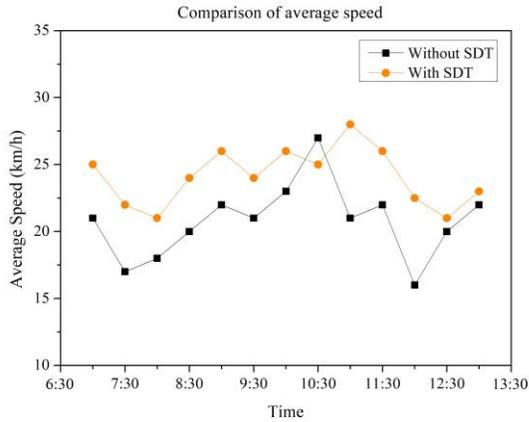

Fig. 14: Verification results

TABLE II: The composition of the Transportation Internet verification system

| Type | Number | Description |
|---|---|---|
| SDT Controller | 1 | Including the edge controller and central controller, running on two Dell servers separately. |
| CAV | 2 | Supporting level4 automated driving and SDV, and connecting to SDT controller through 5G network. |
| Intelligent signal light | 2 | Supporting vehicle-road coordination and software defined transportation infrastructure. |
| Roadside camera and radar | 3 | Supporting vehicle recognition and position perception. |
| Remote driving console | 1 | Console for remote driving application. |
| 5G network | 1 | provided by China Unicom, without edge computing |
| Control center | 1 | Providing the management and control function for the Transportation Internet |

### B. Scale verification, Yuhangtang Road, Hangzhou

For a larger scale verification, we choose Yuhangtang road in Hangzhou to conduct the numerical experiments. We obtain the original traffic data from a taxi company. In addition, combined with the driverless vehicle data in CUHK, Shenzhen campus verification, a total of 300 vehicles are simulated according to different proportions of driverless vehicles based on the cellular automata model. The selected section is 3.38km with 5 signal lights and 6 lanes in both directions. The speed limit is 60km/h and the acquisition interval is 10min (Fig. 15). The simulation model also makes the following assumptions:

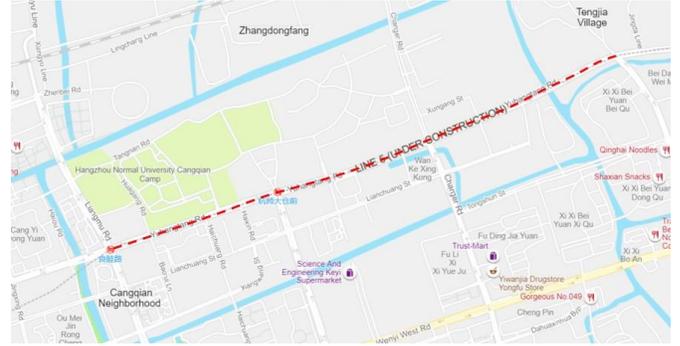

Fig. 15: Verification Site of Yuhangtang Road, Hangzhou

(i) All vehicles are connected to and controlled by SDT controller. The number of driverless vehicles is gradually increased at the proportion of 0%, 20%, 60%.

(ii) Five signal lights are scheduled according to the rules of Green 30s/Yellow 5s/Red 30s. When the vehicle density is less than 10%, the cooperative signal control is provided for connected vehicles with a proportion of 20%.

After the simulation, we can find that the average speed is greatly improved under the control of controller compared with the original data. The average speed is increased by 9.65% without driverless vehicles. When the proportion of driverless vehicles is 20%, the average speed is increased by 13.59%. When the proportion of driverless vehicles is 60%, the average speed is increased by 20.58% (Fig. 16). Because driverless vehicles reduce the headway by platooning, and cooperated with intelligent signal lights, they can further improve the transportation efficiency.
8

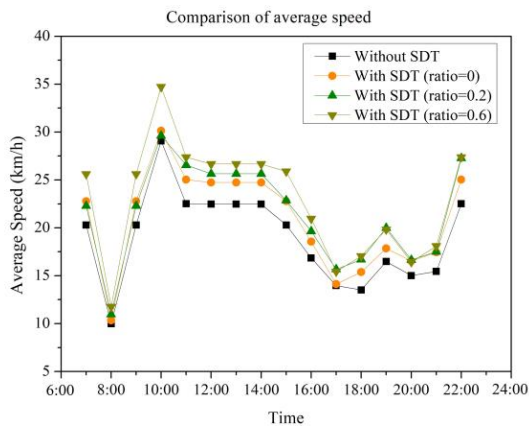

Fig. 16: The comparison of the average speed with SDT

## VII. CONCLUSIONS AND FUTURE WORK

This paper is the first to introduce the Internet paradigm to the transportation field and proposes the Transportation Internet. The Transportation Internet can systematically refer to theories, models, architectures, technologies, algorithms, applications, and business models from the Information Internet and redefine the transportation system. This study is preliminary. A more comprehensive research, including transportation routing algorithms, software-defined architecture, and traffic flow engineering, will be conducted. We believe that the proposed system opens a new field of research and can lay the foundation for the new generation of transportation.